\documentclass[manuscript]{aastex}
\usepackage{spr-astr-addons}

\begin{document}

\slugcomment{draft of \today}

\shorttitle{Cosmological Shock Waves}
\shortauthors{Ryu and Kang}

\title{Shock Waves in the Large-Scale Structure of the Universe}
\author{Dongsu Ryu}
\affil{Department of Astronomy and Space Science,
Chungnam National University, Daejeon 305-764, Korea\\
email: ryu@canopus.cnu.ac.kr}
\and
\author{Hyesung Kang}
\affil{Department of Earth Sciences,
Pusan National University, Pusan 609-735, Korea\\
email: kang@uju.es.pusan.ac.kr}

\begin{abstract}

Cosmological shock waves are induced during hierarchical formation
of large-scale structure in the universe.
Like most astrophysical shocks, they are collisionless,
since they form in the tenuous intergalactic medium through 
electromagnetic viscosities. 
The gravitational energy released during structure formation is
transferred by these shocks to the intergalactic gas as heat,
cosmic-rays, turbulence, and magnetic fields.
Here we briefly  describe the properties and consequences of the shock
waves in the context of the large-scale structure of the universe.

\end{abstract}


\section{Introduction}

Shock waves are ubiquitous in astrophysical environments;
from solar winds to the largest scale of the universe 
\citep{rkhj03}.
In the current paradigm of the cold dark matter (CDM) cosmology,
the large-scale structure of the universe forms through hierarchical
clustering of matter.
Deepening of gravitational potential wells causes gas to move
supersonically.
Cosmological shocks form when the gas accretes onto clusters, filaments,
and sheets, or as a consequence of the chaotic flow motions of the gas
inside the nonlinear structures.
The gravitational energy released during the formation of
large-scale structure in the universe is transferred by these shocks
to the intergalactic medium (IGM).

Cosmological shocks are collisionless shocks which form in a tenuous
plasma via collective electromagnetic interactions between baryonic
particles and electromagnetic fields \citep{quest88}.
They play key roles in governing the nature of the IGM 
through the following processes:
in addition to the entropy generation, cosmic-rays (CRs) are produced
via diffusive shock acceleration (DSA) \citep{bell78,bo78}, magnetic
fields are generated via the Biermann battery mechanism \citep{kcor97}
and Weibel instability \citep{msk06} and also amplified by streaming CRs
\citep{bell04}, and vorticity is generated at curved shocks
\citep{binn74}.

\begin{figure*}[t]
\vspace{0.2cm}
\begin{center}\includegraphics[scale=0.72]{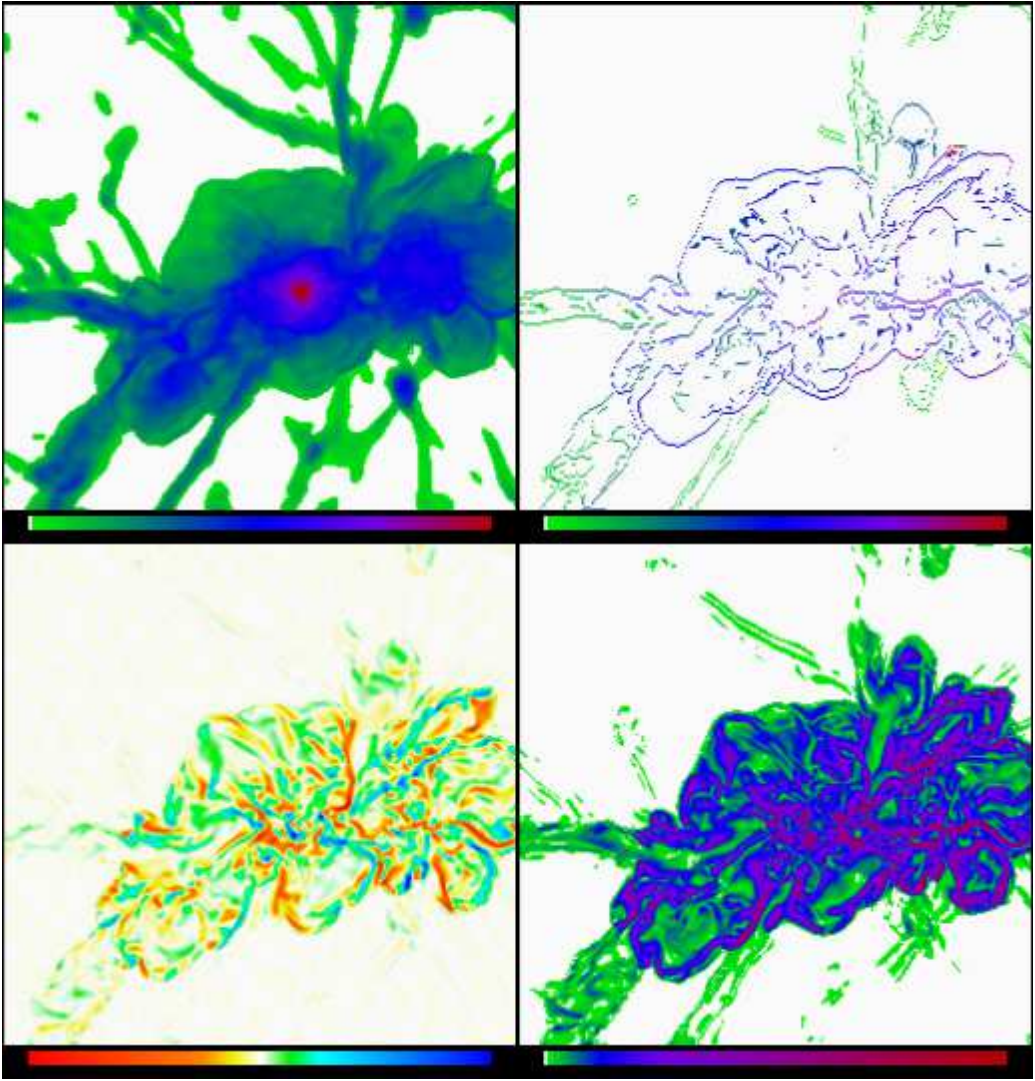}\end{center}
\vspace{0cm}
\caption{Two-dimensional images showing x-ray emissivity (top left),
locations of shocks with color-coded shock speed $V_s$ (top right),
perpendicular component of vorticity (bottom left), and magnitude of
vorticity (bottom right) in the region of $(25\ h^{-1}{\rm Mpc})^2$ 
around a galaxy cluster at present ($z=0$).
Color codes $V_s$ from 15 (green) to 1,800 km s$^{-1}$ (red).}
\end{figure*}

Cosmological shocks in the intergalactic space have been studied in
details using various hydrodynamic simulations for the cold dark matter
cosmology with cosmological constant ($\Lambda$CDM)
\citep{rkhj03,psej06,krco08}.
In this contribution, we describe the properties of cosmological
shocks and their implications for the intergalactic plasma from a
simulation using a PM/Eulerian hydrodynamic cosmology code \citep{rokc93}
with the following parameters:
$\Omega_{BM}=0.043$, $\Omega_{DM}=0.227$, and $\Omega_{\Lambda}=0.73$,
$h \equiv H_0$/(100 km/s/Mpc) = 0.7, and $\sigma_8 = 0.8$.
A cubic region of comoving size 100 $h^{-1}$ Mpc was simulated with
$1024^3$ grid zones for gas and gravity and $512^3$ particles for
dark matter, allowing a uniform spatial resolution of
$\Delta l = 97.7h^{-1}$ kpc.
The simulation is adiabatic in the sense that it does not include
radiative cooling, galaxy/star formation, feedbacks from
galaxies/stars, and reionization of the IGM.
A temperature floor was set to be 
the temperature of cosmic background radiation.
 
\section{Properties of Cosmological Shocks}

As a post-processing step, shocks in the simulated volume are identified 
by a set of criteria based on the shock jump conditions. 
Then the locations and properties of the shocks such as shock speed
($V_s$), Mach number ($M$), and kinetic energy flux ($f_{\rm kin}$) are
calculated.

In the top panels of Figure 1, we compare the locations of
cosmological shocks with the x-ray emissivity in the region
around a cluster of galaxies, both of which are calculated from
the simulation data at redshift $z=0$.
External shocks encompass this complex nonlinear structure 
and define the outermost boundaries up to $\sim 10\ h^{-1}$ Mpc 
from the cluster core,
far beyond the region observable with x-ray of size $\sim 1\ h^{-1}$ Mpc.
Internal shocks are found within the region bounded by external shocks.
External shocks have high Mach numbers of up to $M \sim 10^3$ 
due to the low temperature of the accreting gas in the void region. 
Internal shocks, on the other hand, have mainly low Mach numbers of
$M \la 3$, because the gas inside nonlinear structures has been
previously heated by shocks and so has high temperature. 

\begin{figure}[t]
\vspace{-0.5cm}
\begin{center}\includegraphics[scale=0.415]{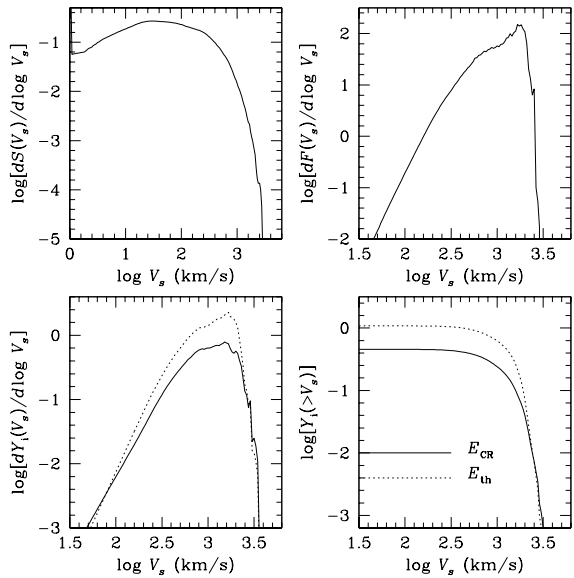}\end{center}
\vspace{0cm}
\caption{(Top left) Reciprocal of the mean comoving distance between
shock surfaces at $z=0$ in units of $1/(h^{-1}{\rm Mpc})$.
(Top right) Kinetic energy flux passing through shock surfaces
per unit comoving volume at $z=0$ in units of $10^{40}$ ergs
s$^{-1}$ $(h^{-1}{\rm Mpc})^{-3}$.
(Bottom left) Thermal energy dissipated (dotted line) and CR energy
generated (solid line) at shock surfaces, integrated from $z=5$ to 0.
(Bottom right) Cumulative energy distributions.
The energies in the bottom panels are normalized to the total thermal
energy at $z=0$.
All quantities are plotted as a function of shock speed $V_s$.}
\end{figure}

The frequency of cosmological shocks in the simulated volume is
represented  by the quantity $S$, the area of shock surfaces per unit
comoving volume, in other words, the reciprocal of the mean comoving
distance between shock surfaces. 
In the top left panel of Figure 2, we show $S(V_s)$ per unit logarithmic
shock speed interval at $z=0$.
We note that the frequency of low speed shocks with $V_s < 15$ km s$^{-1}$ 
is overestimated here, since the temperature of the intergalactic medium is
unrealistically low in the adiabatic simulation without the cosmological 
reionization process. 
Although shocks with $V_s \sim$ a few $\times$ 10 km s$^{-1}$ are most
common, those with speed up to several $\times 10^3$ km s$^{-1}$ are
present at $z=0$.
The mean comoving distance between shock surfaces is
$1/S \sim 3\ h^{-1}{\rm Mpc}$ when averaged over the entire universe, 
while it is $\sim 1\ h^{-1}{\rm Mpc}$ inside the nonlinear structures
of clusters, filaments, and sheets.

In order to evaluate the energetics of cosmological shocks, the incident
shock kinetic energy flux, $f_{\rm kin} = (1/2) \rho_1 V_{s}^3$, is
calculated.
Here $\rho_1$ is the preshock gas density.
Then the average kinetic energy flux through shock surfaces per unit
comoving volume, $F$, is calculated. 
The top right panel of Figure 2 shows $F(V_s)$ per unit logarithmic
shock speed interval.
Energetically the shocks with $V_s > 10^3$ km s$^{-1}$, which
form in the deepest gravitation potential wells in and around clusters
of galaxies, are most important.
Those responsible for most of shock energetics are the internal shocks
with relatively low Mach number of $M \sim 2 - 4$ in the hot IGM,
because they form in the high-density gas inside nonlinear structures.
On the other hand, external shocks typically form in accretion flows
with the low-density gas in voids, so the amount of the kinetic energy
passed  through the external shocks is rather small.

\section{Energy Dissipation at Cosmological Shocks}

\begin{figure}[t]
\vspace{-2.5cm}
\begin{center}\includegraphics[scale=0.415]{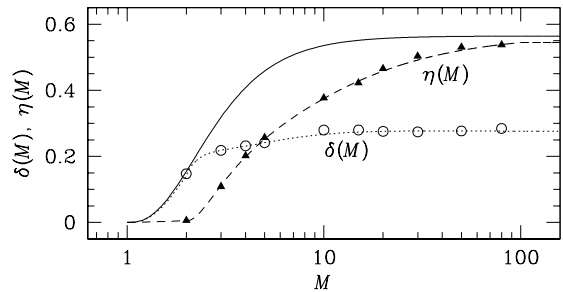}\end{center}
\vspace{-2cm}
\caption{Gas thermalization efficiency, $\delta(M)$, and CR
acceleration efficiency, $\eta(M)$, as a function of Mach number.
Symbols are the values estimated from numerical simulations based on
a DSA model and dotted and dashed lines are the fits.
Solid line is for the gas thermalization efficiency for shocks
without CRs.}
\end{figure}

In addition to the gas entropy generation, the acceleration of CRs
is an integral part of collisionless shocks, in which electromagnetic
interactions between plasma and magnetic fields provide the necessary 
viscosities.
Supra-thermal particles are extracted from the shock-heated thermal
particle distribution \citep{md01}.
With $10^{-4} - 10^{-3}$ of the particle flux passing through the shocks
injected into the CR population, up to $\sim$60\% of the kinetic
energy of strong quasi-parallel shocks can be converted into CR ions
and the nonlinear feedback to the underlying flow can be substantial
\citep{kj05}.
At perpendicular shocks, however, the CR injection and acceleration are
expected to be much less efficient, compared to parallel shocks, since
the transport of low energy particles normal to the average field
direction is suppressed.
So the CR acceleration depends sensitively on the mean magnetic field
orientation.

Time-dependent simulations of DSA at quasi-parallel shocks with a thermal
leakage injection model and a Bohm-type diffusion coefficient have shown
that the evolution of CR modified shocks becomes self-similar, after 
the particles are accelerated to relativistic energies and the precursor
compression reaches a time-asymptotic state \citep{kj05,kj07}.
The self-similar evolution of CR modified shocks depends somewhat 
weakly on the details of various particle-wave interactions, but 
it is mainly determined by the shock Mach number. 
Based on this self-similar evolution, we can estimate the gas 
thermalization efficiency, $\delta(M)$, and the CR acceleration efficiency,
$\eta(M)$, as a function shock Mach number, which represent the fractions
of the shock kinetic energy transferred into the thermal and CR energies,
respectively. 
Figure 3 shows the results of such DSA simulations \citep{krco08}.
From the figure, we expect that at weak shocks with $M \la 3$ 
the energy transfer to CRs should be $\la 10$ \% of the shock kinetic
energy at each shock passage, whereas at strong shocks with $M \ga 3$
the transfer is very efficient and the flow should be significantly
modified by the CR pressure.

Adopting the efficiencies shown in Figure 3, the thermal and CR energy
fluxes dissipated at each cosmological shock can be estimated as
$\delta(M) f_{kin}$ and $\eta(M) f_{kin}$, respectively.
Then the total energies dissipated through the surfaces of cosmological
shocks during the large-scale structure formation of the universe
can be calculated \citep{rkhj03,krco08}.
The bottom panels of Figure 2 show the thermal and CR energies,
integrated from from $z=5$ to 0, normalized to the total gas thermal
energy at $z=0$, as a function of shock speed.
Here we assume CRs are freshly injected at shocks without a
pre-existing population.
The shocks with $V_s>10^3$ km s$^{-1}$ are most
responsible for the shock dissipation into heat and CRs.
The figure shows that the shock dissipation can count most of the gas
thermal energy in the IGM \citep{krcs05}.
The ratio of the total CR energy, $Y_{CR}(>V_{s, {\rm min}}$), to 
the total gas thermal energy, $Y_{th}(>V_{s,{\rm min}}$), dissipated at
cosmological shocks throughout the history of the universe is about 0.4
(where $V_{s, {\rm min}}$ is the minimum shock speed), giving a rough
estimate for the energy density of CR protons relative to that of thermal
gas in the IGM as $\varepsilon_{\rm CRp} \sim 0.4 \varepsilon_{\rm therm}$.

Because of uncertainties in the DSA model such as the field obliquity and
the injection efficiency, however, it is not meant to be an accurate
estimate of the CR energy in the IGM.
Yet the results imply that the IGM could contain a dynamically
significant CR population.

\section{Turbulence Induced by Cosmological Shocks}

\begin{figure}[t]
\vspace{0cm}
\begin{center}\includegraphics[scale=0.415]{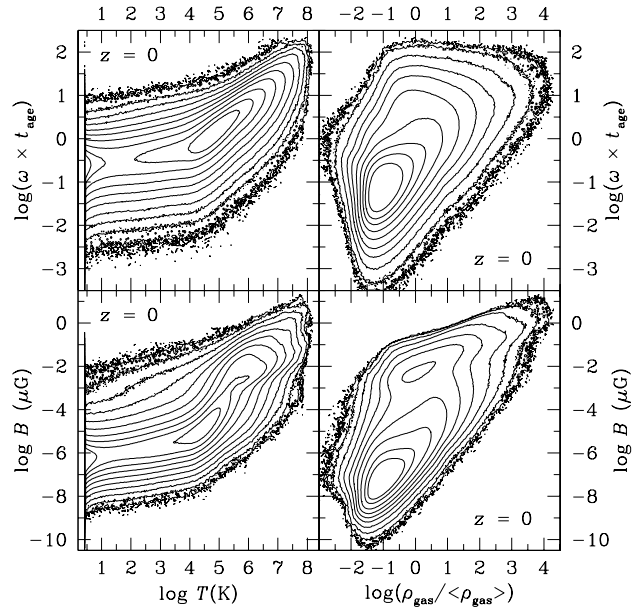}\end{center}
\vspace{0cm}
\caption{Volume fraction with given temperature and vorticity magnitude
(top left), with given gas density and vorticity magnitude (top right),
with given temperature and magnetic field strength (bottom left), and
with given gas density and magnetic field strength (bottom right)
at present ($z=0$).
Here, $t_{\rm age}$ is the present age of the universe.}
\end{figure}

Vorticity can be generated in the IGM either directly at curved
cosmological shocks or by the baroclinity of flow. 
The baroclinity is resulted mostly from the entropy variation induced
at cosmological shocks.
Therefore, the baroclinic vorticity generation also can be attributed
to the presence of cosmological shocks.
A quantitative estimation of the vorticity in the IGM was made
using the data of the simulation described in \S 1 \citep{rkcd08}.
The bottom panels of Figure 1 show the distribution of the vorticity
around a cluster complex.
The distribution closely matches that of shocks, as expected.

The top panels of Figure 4 show the magnitude of the vorticity 
in the simulated volume as a function of gas temperature and density.
Here the vorticity magnitude is given in units of the reciprocal 
of the age of the universe, $1/t_{\rm age}$.
There is a clear trend that the vorticity is larger in hotter and
denser regions.
At the present epoch, the rms vorticity is
$\omega_{\rm rms}t_{\rm age} \sim 10$ to $30$ in the regions 
associated with clusters/groups of galaxies ($T > 10^7$ K) 
and filaments ($10^5 < T < 10^7$ K), whereas it is on
the order of unity in sheetlike structures with $10^4 < T < 10^5$ K
and even smaller in voids with $T < 10^4$ K.

With $t_{\rm eddy} = 1/\omega$ interpreted as the local eddy turnover
time, $\omega \times t_{\rm age}$ represents the number of eddy
turnovers of vorticity in the age of the universe.
It takes a few turnovers for vorticity to decay into smaller eddies 
and develop into turbulence.
So with $\omega_{\rm rms}t_{\rm age} \sim 10 - 30$, the flows in
clusters/groups and filaments is likely to be in the state of turbulence.
On the other hand, with $\omega_{\rm rms}t_{\rm age} \la 1$ the flow in
sheetlike structures and voids is expected to be mostly non-turbulent.

In order to estimate the energy associated with the turbulence in the IGM, 
the curl component of flow motions, ${\vec v}_{\rm curl}$, which
satisfies the relation
${\vec\nabla}\times{\vec v}_{\rm curl} \equiv {\vec\nabla}\times{\vec v}$,
is extracted from the velocity field.
As vorticity cascades and develops into turbulence, the energy
$(1/2)\rho v_{\rm curl}^2$ is transferred to turbulent motions, so
it can be regarded as the turbulence energy, $\varepsilon_{\rm turb}$.
As shown in \citet{rkcd08},
$\varepsilon_{\rm turb} < \varepsilon_{\rm therm}$ in clusters/groups.
In particular, the mass-averaged value is $\left<\varepsilon_{\rm turb}
/\varepsilon_{\rm therm}\right>_{\rm mass} = 0.1-0.3$ in the
intracluster medium (ICM), which is in good agreement with the
observationally inferred values in cluster core regions \citep{sfmb04}.
In filaments and sheets, this ratio is estimated to be 
$0.5 \la \left<\varepsilon_{\rm turb}/ \varepsilon_{\rm therm}\right> \la 2$
and it increases with decreasing temperature.

\section{Intergalactic Magnetic Field}

How have the intergalactic magnetic fields (IGMFs) arisen?
The general consensus is that there was no viable mechanism to produce
strong, coherent magnetic fields in the IGM prior to the formation
of large-scale structure and galaxies \citep{kz07}.
However, it is reasonable to assume that weak seed fields were
created in the early universe.  A number of mechanisms, including
the Biermann battery mechanism \citep{kcor97} and Weibel instability
\citep{msk06} working at early cosmological shocks, have been suggested
\citep{kz07}.
The turbulence described in \S 4 then can amplify the seed fields in
the IGM through the stretching of field lines, a process known as
the turbulence dynamo.
In this scenario the evolution of the IGMFs should go through three stages:
(i) the initial exponential growth stage, when the back-reaction of
magnetic fields is negligible;
(ii) the linear growth stage, when the back-reaction starts to operate;
and (iii) the final saturation stage \citep{cv00,choetal08}.

In order to estimate the strength of the IGMFs resulted from the dynamo
action of turbulence in the IGM, we model the growth and saturation of
magnetic energy as: 
\begin{eqnarray}
\frac{\varepsilon_B}{\varepsilon_{\rm turb}}
= \left\{ \begin{array}{ll}
0.04 \times \exp\left[(t'-4)/0.36\right]&{\rm for}\ t'<4, \\
(0.36/41) \times (t'-4) + 0.04&{\rm for}\ 4<t'<45, \\
0.4 &{\rm for}\ t'>45,
\end{array} \right.
\nonumber
\end{eqnarray}
based on a simulation of incompressible magnetohydrodynamic turbulence
\citep{rkcd08,choetal08}.
Here $t' = t/t_{\rm eddy}$ is the number of eddy turnovers.
This provides a functional fit for the fraction of the turbulence energy,
$\varepsilon_{\rm turb}$, transfered to the magnetic energy,
$\varepsilon_B$, as a result of the turbulence dynamo.

The above formula is convoluted to the data of the simulation described
in \S 1, setting $t' \equiv \omega \times t_{\rm age}$, and the strength
of the IGMFs is calculated as $B=(8\pi\varepsilon_{\rm B})^{1/2}$.
The resulting magnetic field strength is presented in the bottom panels
of Figure 4.
On average the IGMFs are stronger in hotter and denser regions.
The strength of the IGMFs is $B \ga 1 \mu$G inside clusters/groups
(the mass-averaged value for $T > 10^7$ K), $\sim 0.1 \mu$G around
clusters/groups (the volume-averaged value for $T > 10^7$ K), and
$\sim 10$ nG in filaments (with $10^5 < T < 10^7$ K) at present.
The IGMFs should be much weaker in sheetlike structures and voids.
But as noted above, turbulence has not developed fully in such low
density regions, so our model is not adequate to 
predict the field strength there.

We note that in addition to the turbulence dynamo, other processes such
as galactic winds driven by supernova explosions and jets from active
galactic nuclei can further strengthen the magnetic fields to the IGM
\citep[for references, see][]{rkcd08}

\section{Conclusion}
Shocks are inevitable consequences of the formation of large-scale
structure in the universe.
They heat gas, accelerate cosmic-ray particles, produce vorticity and
turbulence, and generate and amplify magnetic fields in the IGM.
By applying detailed models of the DSA and the turbulence dynamo to the
data of a cosmological hydrodynamic simulation of a concordance
$\Lambda$CDM universe, we have made the following quantitative estimates: 
$\varepsilon_{\rm CRp} / \varepsilon_{\rm therm} \sim 0.4$ in the IGM,
$\left<\varepsilon_{\rm turb} /\varepsilon_{\rm therm}\right>_{\rm mass}
= 0.1 - 0.3$ in the ICM,
$\left<\varepsilon_{\rm turb}/ \varepsilon_{\rm therm}\right> \sim 0.5 - 2$
in filaments and sheets,
$B \ga 1 \mu$G inside clusters/groups, $B \sim 0.1 \mu$G around
clusters/groups, and $B \sim 10$ nG in filaments at present.
Our results suggest that the non-thermal components can be energetically
significant  in the intergalactic plasma of large-scale structure, 
as in the interstellar plasma inside Our Galaxy. 

\acknowledgments
The work of DR was supported by the Korea Research Foundation Grant
funded by the Korean Government (MOEHRD) (KRF-2007-341-C00020).
The work of HK was supported for two years by Pusan National University
Research Grant.

\end{document}